\documentstyle[12pt,epsf]{article}
\def\beq{\begin{equation}}   \def\eeq{\end{equation}}

\begin{document}
\begin{flushright}
NYU-TH/99/08/01 \\
TPI-MINN-99/38-T\\
UMN-TH-1813/99 \\
hep-th/9910050\\ 
\end{flushright}

\vspace{0.1in}
\begin{center}
\bigskip\bigskip
{\large \bf D-Walls and Junctions in Supersymmetric Gluodynamics
\\ \vspace{0.1in} 
in the Large
$N$  Limit
\\ \vspace{0.1in} 
Suggest the Existence of  Heavy Hadrons}
\vspace{0.3in}      

{Gregory Gabadadze$^1$,~~Mikhail Shifman$^2$}
\vspace{0.1in}

{\baselineskip=14pt \it 
$^1$Department of Physics, New York University, 4 Washington Place, 
New York, NY 10003 } \\
{\baselineskip=14pt \it 
$^2$Theoretical Physics Institute, University of Minnesota, Minneapolis, 
MN 55455} \\
\vspace{0.2in}
\end{center}

\vspace{0.9cm}
\begin{center}
{\bf Abstract}
\end{center} 
\vspace{0.1in}

A number of arguments exists that the ``minimal"  BPS wall 
width in large $N$ supersymmetric gluodynamics
vanishes  as $1/N$.  There  is  a certain tension  between this assertion  
and the fact that the mesons coupled to $\lambda\lambda$
have masses ${\cal O}(N^0)$.  To reconcile these facts  
we argue that there should exist additional
soliton-like states  with  masses   scaling as
$N$.  The  BPS walls  must be  ``made" predominantly of these  
heavy states which are coupled to $\lambda \lambda$ stronger than the 
conventional mesons.  The tension 
of the BPS wall  junction
 scales  as $N^2$, which serves  as an additional argument
in favor of the $1/N$ scaling of the wall width. 
The heavy states can be thought of as  solitons
of the corresponding closed string theory.
They  are related to  certain fivebranes   
in the M-theory construction.
We study  the issue of the wall width in 
toy models which capture  some features 
of supersymmetric gluodynamics.  We speculate
that the special hadrons with mass scaling as $N$ should also exist 
in the large $N$ limit of {\em non-supersymmetric} gluodynamics.

\newpage
\section{Introduction}

Supersymmetric gluodynamics is the simplest non-Abelian 
supersymmetric (SUSY)
gauge theory which turns out hardest to handle. 
The only global symmetry in this model (besides SUSY itself)
is ${\bf Z}_{2N}$ discrete symmetry which is spontaneously broken
by the gluino condensate down to ${\bf Z}_2$ (the gauge group
here is assumed to be SU$(N)$). The lack of a continuous
moduli space of vacua  with a rich set of global 
symmetries prevents one from studying the theory 
in the  spirit of Seiberg \cite {Seiberg}.

Possible  existence of the BPS domain walls in SUSY 
gluodynamics \cite 
{DvaliShifman,KovnerSmilgaShifman,ChibisovShifman} 
adds a significant, albeit indirect, information to a rather fragmentary
knowledge of the strong coupling dynamics of this theory. 
What is the most surprising property of the BPS walls?

It was argued in \cite {DvaliGabadKakush} that
the width $l$ of the ``minimal" wall
(i.e. the wall connecting the neighboring vacua)  must scale as $1/N$ in
the large
$N$ limit. By the width we mean the transverse dimension saturating 
the
wall tension. At first sight it is natural to expect this dimension to be of
order 1 rather than $1/N$. Indeed, the order parameter 
$\lambda\lambda$  is a source of particles
with mass ${\cal O}(1)$, the ``gluinoballs."
This means that at large separations from the wall,
$\Delta z\gg \Lambda^{-1}$, the wall tails should fall off as
${\rm exp} (-~\Lambda \Delta z)$, implying that the wall
width is ${\cal O}(1)$. (Here and below the scale parameter is denoted 
by $\Lambda$.)
 
 The conclusion that $l\sim 1/N$ was further supported 
in Ref. \cite {DvaliGabadKakush}  by constructing the BPS walls 
in a certain model for SUSY gluodynamics. 
One may wonder to which extent this construction is 
model-independent. We first summarize  evidence in favor of 
$l\sim 1/N$. A new argument comes from a rather
unexpected side.  It was pointed out recently that  
several walls can join together and 
form a wall junction configuration preserving
$1/4$ of the original SUSY \cite {Townsend,Carroll} (see Fig. 1). 
The models considered in the literature previously
are those of the Wess-Zumino type. We will assume
that the BPS junctions with $N$ minimal  walls exist in
SU($N$) supersymmetric gluodynamics. Then, the $N$ dependence of the
junction tension comes out natural
if $l$ is assumed to be ${\cal O} (1/N)$; on the contrary,
it is very hard to get the proper junction tension under any other scaling
law for $l$. 

Given that the BPS wall width $l\sim 1/N$ one can raise a legitimate 
question
how this can possibly be explained in terms of 
the physical excitations of the theory.
We will  argue  that 
there should exist certain distinguished
(soliton-like) excitations in the theory 
with mass $m\sim {\cal O} (N)$. These excitations are coupled
to $\lambda\lambda$ stronger that the conventional mesons, so that
the wall is ``made" predominantly of these heavy states. 
Some arguments in favor of the existence of such states
can be found within the  D-brane construction 
of SUSY gluodynamics due to Witten \cite {WittenMQCD}, which also 
leads to the  conclusion, that $l \sim 1/N$. We believe that
this conclusion is general -- as long as the  minimal wall is
BPS saturated, its width has to be ${\cal O} (N^{-1})$,
which entails, in turn, that matter of which it is (predominantly) built
 has mass ${\cal O} (N)$. Various simple toy models
illustrate this assertion. 

If one introduces the gluino mass, supersymmetry is broken.
At small masses the breaking is small -- we are pretty close to the
supersymmetric picture. Once the gluino mass becomes larger than 
$\Lambda$,
the gluinos decouple, and we find ourselves in 
non-supersymmetric  Yang-Mills
theory. One can try to extrapolate in the gluino mass, assuming that
there is no phase transition. Speculating along these lines, we
have to conclude that the $M\sim N$ hadrons of a  special
nature must exist in non-supersymmetric gluodynamics too.
This conclusion can also be supported by the D-brane construction in
non-supersymmetric Yang-Mills theory where  domain walls 
can  be continuously extrapolated to a certain wrapped D-brane.  

The paper is organized as follows. In
Sec. 2 we briefly summarize what is known about the minimal
BPS wall width in SUSY gluodynamics. The $N$ counting for the BPS 
junction
is discussed. In Sec. 3 we give a field-theoretic argument
in favor of  $M\propto N$ (here $M$ is the mass of the quantum of 
which
the wall is ``built").   In Sec. 4  we identify possible (soliton-like)
candidates for the ``building blocks" of the  minimal walls within the 
D-brane construction.  
In Sec. 5 we introduce a (SUSY violating) gluino mass, and perform
the extrapolation to large masses. This gives us a hint of
the existence of the $M\sim N$ hadrons in the large $N$ limit of non-
supersymmetric
gluodynamics. We ameliorate these arguments by an evidence coming 
from the 
D-brane construction.
Finally, 
Sec. 6 is devoted to toy models.
The BPS wall junctions 
are studied in this section as well.

\begin{figure}
\centerline{\epsfbox{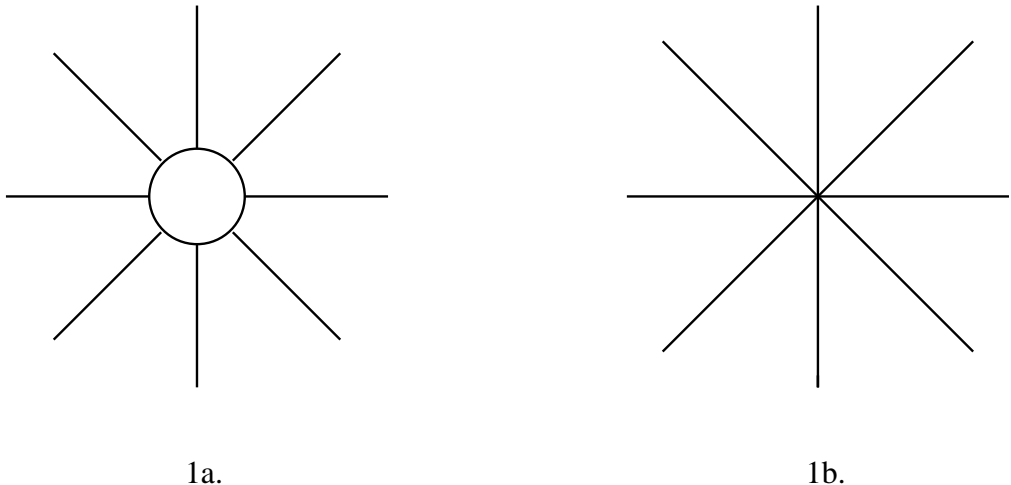}}
\epsfysize=6cm
\caption{\small Domain wall junctions}
\label{fig1}
\end{figure} 

\section{Why the minimal wall width must scale as $1/N$}

There are at least three different arguments supporting 
this conclusion. First, as is well-known, a natural behavior of the 
volume energy density inside the wall is
\footnote{More accurately, one should speak of the difference between 
the
volume energy density inside the wall and in the vacuum. The 
latter vanishes, however,
in supersymmetric theories.}
$\varepsilon \sim N^2$ (since there are $N^2$ degrees of freedom in the 
theory). Then, the fact that the BPS wall tension is 
$\sim N$ \cite {DvaliShifman} implies that $l\sim 1/N$
\cite{DvaliGabadKakush}.  This conclusion is supported  by the solutions 
found
in \cite {DvaliKakush} and \cite {DvaliGabadKakush}. 

The second argument is  based on  
the D-brane construction~\cite {WittenMQCD}, which  
also implies that  $l \sim 1/N$ (see Sec. 4 for further details). 

Here we will focus on a new argument based on the wall junction \cite
{Townsend,Carroll}. We will assume that the BPS saturated junction of 
$N$
minimal walls exists in supersymmetric gluodynamics. Then, 
as was shown in Ref. \cite{GorskiShifman}, the junction tension 
$T_{\rm Junction}$
is
\begin{equation}
T_{\rm Junction} \propto \oint a_k \, dx^k \,  \propto\, N^2\, .
\label{dopone}
\end{equation}
Here $a_\mu$ is the axial current of gluinos (which scales as
$N^2$), and the integration contour runs over the large circle in the 
plane
perpendicular to the wall junction. Let us try to understand 
geometrically how this behavior could occur.

Let us examine the ``spokes" and the ``hub" in Fig. 1a. If the width of the
spokes is $1/N$, and there are $N$ of them, then the diameter
of the hub is ${\cal O}(N^0)$. The area of the hub is ${\cal O}(N^0)$
too. Given that the volume energy density scales as $N^2$,
we naturally arrive at $T_{\rm Junction} \propto N^2$.

At the same time, if the width of the
spokes is ${\cal O}(N^0)$, as was naively believed before 
\cite{DvaliGabadKakush}, the diameter
of the hub is ${\cal O}(N)$ while the area of the hub is ${\cal O}(N^2)$.
To explain Eq. (\ref{dopone}) one must assume that the volume energy
density inside the hub differs from the vacuum energy density
by ${\cal O}(N^0)$, which is extremely counterintuitive. 
 
Although, the arguments listed above are qualitative and 
present only 
circumstantial  evidence, taken together they seem compelling. 
We will accept
that the BPS wall width $l\sim 1/N$ as a  starting point.
The conclusion of the existence of heavy soliton-like
hadrons with mass scaling as $\Lambda N$ will ensue
\cite{AVa}.

 \section{Suggestive  Arguments from  Field-Theoretic \\
Consideration}

Supersymmetric gluodynamics is a strongly coupled theory.
This means that we have no reliable tools for addressing
the problem we are interested in at the quantitative level.
However,  we can consider the issue at the qualitative
level if, instead of SUSY gluodynamics, we will deal with a simpler 
theory
belonging to the same universality class.
The most natural choice seems to be SQCD with
$N_f = N-1$. If the matter mass term is small, $m\to 0$,
this is a weakly coupled theory (in the Higgs phase).
 It explicitly exhibits the
${\bf Z}_N$ structure of the vacuum state. The minimal BPS walls
can be treated quasiclassically. We will show that 
their width shrinks
with $N$ as $1/N$,
while the mass of the quantum they are built of grows linearly with $N$.

At the same time, since all matter fields are in the fundamental
representation, it seems reasonable to assume that there is no phase
transition that would separate the Higgs phase at small
$m$ from the strongly coupled phase at large $m$.
At large $m$ the matter fields decouple, and we return to 
SUSY gluodynamics. It is natural to think that the
interpolation  is smooth in this transition.
The field with mass $\propto N$ comprising the wall
at small coupling goes  into a gluon/gluino ``soliton"
with mass $M\sim N\Lambda$ in SUSY gluodynamics.

In more detail, the construction is as follows.
One introduces $N_f$ chiral superfields $Q_f$ ($f=1,2,...,N_f$) which are
fundamentals of SU($N$) and $N_f$ chiral superfields $\tilde Q^g$
($g=1,2,...,N_f$) which are antifundamentals. Here $N_f = N -1$.
We then introduce the tree level superpotential
\begin{equation}
{\cal W}_{\rm tree} = m\sum_{f=1}^{N_f} M_f^f~,
\label{mt}
\end{equation}
where
\begin{equation}
M_f^g \equiv Q_f \tilde Q^g
\label{igsp}
\end{equation}
are the moduli, and the mass term $m$ is assumed to be very small and 
diagonal.
Then the theory is fully Higgsed \cite{ADS}. A superpotential is 
generated on the
moduli space (via instantons) once the   heavy fields are integrated out,
\begin{equation}
{\cal W}_{\rm inst} = \frac{\tilde\Lambda^{2N+1}}{{\rm det} M}\, ,
\label{igspm}
\end{equation}
where $\tilde\Lambda$ is a scale parameter of SQCD (it
will drop out in what follows). 

If the mass term is chosen as in Eq. (\ref{mt}),
the vacuum expectation values of the moduli are diagonal too,
\begin{equation}
\langle M_f^g \rangle = v^2 \delta_f^g\, .
\label{vevsp}
\end{equation}
There are $N$ solutions for $v^2$,
$$
v^2 = |v^2| \exp \left(\frac{2\pi i k}{N}\right)\,,\quad k = 1,2, ... , N.
$$
The absolute value of $v^2$
is very large in the limit of small $m$.
This explains why the vector bosons and their superpartners are 
heavy and can be
integrated out. Equation (\ref{vevsp}) implies that
although there are $N_f^2$ light moduli fields,
the domain walls occur only for the field $\sum_{f=1}^{N_f} M_f^f$.
All other $N_f^2-1$ moduli fields, say $M_1^2$ or $M_1^1 - M_2^2$, etc.,
oscillate near zero; they do not experience the wall-type transition.

Next, we combine Eqs. (\ref{mt}) and (\ref{igspm}), find the 
stationary points of the
superpotential to determine the vacua, and then find the masses
of all $N_f^2$ fields on the moduli space.
The mass of the quantum for which $\sum_{f=1}^{N_f} M_f^f$ is the 
interpolating 
field is $Nm$; all other mass eigenvalues 
do not contain  the $N$ factor.

The minimal BPS  walls in this model were considered in Ref. \cite{AVS}.
Although the $N$ dependence of the wall width was not
explicitly discussed, one can infer from these works
that the wall is built of the field
$\sum_{f=1}^{N_f} M_f^f$, and its width is $(mN)^{-1}$.
Although other $N_f^2 - 1$  ``genuinely light" moduli do couple to the 
field
$\sum_{f=1}^{N_f} M_f^f$ through Eq. (\ref{igspm}),  they 
are not excited on the
wall. This explains why there is no wall broadening. The same
 phenomenon will be discussed in details
in Sec. 6.

To avoid confusion, a remark is in order here
concerning the relation between the  argument above and the results
of Ref. \cite{AVS}. Smilga and collaborators did find a phase transition in 
the matter mass parameter $m$. This is explained by the fact that they 
were forced to deal with the 
 Taylor-Veneziano-Yankielowicz model on the large $m$ side,
rather than   with  genuine SQCD.
In the  description  of walls this model  is not adequate 
(in fact, it is inadequate both  in the small and  in the large $m$ 
domains).
If $m$ is small, and one retains only moduli, the corresponding 
effective theory is an exact low-energy expansion, in the Wilsonean 
sense.
As such, it describes the dynamics of the moduli in full.
Thus, when $m$ is small, (more exactly, $Nm$ has to be small) and  the 
wall 
profile is broad,  it is fully legitimate to  use the effective theory of 
the moduli to {\em  exhaustively}
describe the walls.  As $Nm$ approaches $\tilde\Lambda$, 
the description of the walls in the 
effective
theory of moduli becomes wrong. At this point we must  roll over from 
the theory
of moduli to  full SQCD.  Then, there will be no phase 
transition in $m$. Thus, the observation of this phase transition in 
Ref. \cite{AVS} is the artifact of the approximation used. 

\section{Suggestive  Arguments from  D-branes}

The aim of this section is to use the D-brane 
construction of the  BPS domain wall \cite {WittenMQCD}
to study the origin of the 
states with the mass of order ${\cal O} (N)$. These  states  should  
be responsible for the $1/N$ scaling
of the domain wall width.  

In the large $N$ limit SUSY gluodynamics is 
expected to be  described by
a certain  non-critical closed string theory (let us call it tentatively the
``closed QCD string"). This theory is 
hard to formulate precisely. However, it is  believed to lie  
in the  universality class of the theory 
obtained via D-brane construction
\cite {WittenMQCD}.
Excitations of the closed QCD string should give rise
to the colorless bound state  spectrum of SUSY gluodynamics. 
These are gluino-gluino, gluino-gluon and pure gluonic
bound states (including  higher radial excitations). 
All these states  have masses  of  order ${\cal O} (1)$. Moreover, 
they couple to each other with couplings  suppressed by
powers of $1/N$. This exhausts the perturbative 
part of the closed QCD string theory. 
In addition, there should exist
a nonperturbative sector of the closed QCD string theory.
In analogy with type IIA, B strings, one might expect  to find  point-like 
or extended objects in the nonperturbative sector
\footnote{ In the given context the reference to point-like objects
means that their size is $1/N$.}. 
These objects have little to do with mesons and glueballs --
they represent independent degrees of freedom
of the theory. 

As a particular consequence of this fact, the BPS domain walls
in  SUSY gluodynamics manifest themselves as certain 
D-branes, on which the  open QCD strings can end \cite {WittenMQCD}. 
Notice, that (naively) the  open 
strings are not expected in SUSY Yang-Mills (SYM) theory, since  
in this theory  
there are no fundamentals 
on which the open strings could  end. 
Nevertheless, the open 
strings appear at nonperturbative level, precisely as it happens 
in type IIA, B critical closed string theories. 
As was mentioned,  
these open strings  terminate on the
D-branes, i.e., on the BPS domain walls  of the underlying SYM
theory 
\cite {WittenMQCD} (see Fig. 2).  
For these reasons, in what follows let us call these BPS 
walls D-walls. It seems clear that
one could not find a D-wall in the theory where only  mesons with
 masses ${\cal O}(1)$  are included.

The effective meson theory
for  the large $N$ gluodynamics is believed to be  
the theory of a closed QCD string (as opposed to the 
open string theory of  conventional QCD with quarks). 
What  kind of  new nonperturbative point-like objects exist 
in the closed string theory?
A point-like soliton of the closed string
theory with the mass of  order ${\cal O}(N)$ reminds   
a zero-brane. Unfortunately, the question 
whether the QCD string has  zero-branes is 
hard to study in the fundamental theory (since no 
consistent string theory is known for this).
What we could study  instead is the theory which is in the same 
universality 
class, and, which can be realized in a particular D-brane construction 
\cite {WittenMQCD}. 
We will argue below that 
the brane construction calls for the inclusion of the
states with mass ${\cal O}(N)$ in order to be able to
describe  the BPS domain walls.  

Consider two parallel BPS walls at a certain distance from each other.
One can stretch between them one, two, and so on,
open QCD strings. At the point of each string-wall junction
one observes a lump of energy which corresponds to an object
similar to a quark in the fundamental representation.
This lump of energy is localized on the wall. 
When the number of the QCD strings stretched
between the walls becomes equal to $N$, the lumps can fuse
forming an object which is color singlet, has mass $m\sim N$ and 
reminds a baryon.  Since it is color singlet, there are no reasons
why it should be localized on the wall -- it seems likely 
that it will propagate in the
bulk.

Thus, we come to a conclusion that there
 might exist some 
new states in SYM theory which are  neither mesons nor glueballs;
nevertheless,  they  couple  to the BPS domain wall. 

In the field-theoretic language one may recall
 that the baryons 
in conventional QCD emerge as the Skyrme solitons in the effective 
meson theory
\cite {Skyrme}. Their  masses  scale  as ${\cal O}(N)$ 
in the large $N$ limit, although the masses of the original mesons
are ${\cal O}(1)  $
\cite {WittenN}. 

To reiterate the argument at a slightly more
quantitative level we first briefly review the  construction of Ref. \cite
{WittenMQCD}. Following \cite {Kutasov,HannanyWitten,WittenMQCD} we 
start
with  type IIA string theory. The brane setup is as follows.
There is one Neveu-Schwarz (NS) fivebrane which spans the 
worldvolume
$(x_0,~x_1,~x_2,~x_3,~x_4,~x_5)$ and which lives at the 
point $x_6=x_7=x_8=x_9=0$. 
Another fivebrane (called NS$^\prime$) with the worldvolume
$(x_0,~x_1,~x_2,~x_3,~x_7,~x_8)$ is separated form
the NS fivebrane at some distance $S_0$ along $x_6$ ($x_6=S_0$).
There are $N$ coincident D4-branes suspended
between the fivebranes (see Fig. 3). The worldvolume
of these branes is in the $(x_0,~x_1,~x_2,~x_3,~x_6)$ 
part of space-time. 
One defines $v=x_4+ix_5,~w=x_7+ix_8$. 
This configuration 
is argued to describe  ${\cal N}=1$ 
SYM theory  without chiral multiplets at low energies. In the infrared 
limit this field theory lives
in the $(x_0,x_1,x_2,x_3)$ subspace.  
The gauge coupling of this 
model is related to the string coupling constant, $g_s$, as follows:
$g_{\rm YM}^2\sim g_{s}l_s/S_0$, where $l_s$ is 
the type IIA string length. 
The model is studied by 
elevating the type IIA construction to M-theory. Thus, one takes
$g_{s}$ to be large so that  the eleventh dimension
opens up \footnote{
Below, we will see that in the particular case 
at hand the coupling which effectively defines
the transition to M-theory is $Ng_s$ instead of $g_s$.
This is related to a membrane which is wrapped $N$ times around the 
compact eleventh dimension.} 
in the form of a circle ${\bf S}^1$ of radius $R$,
\begin{figure}[h]
\begin{center}
\epsfbox{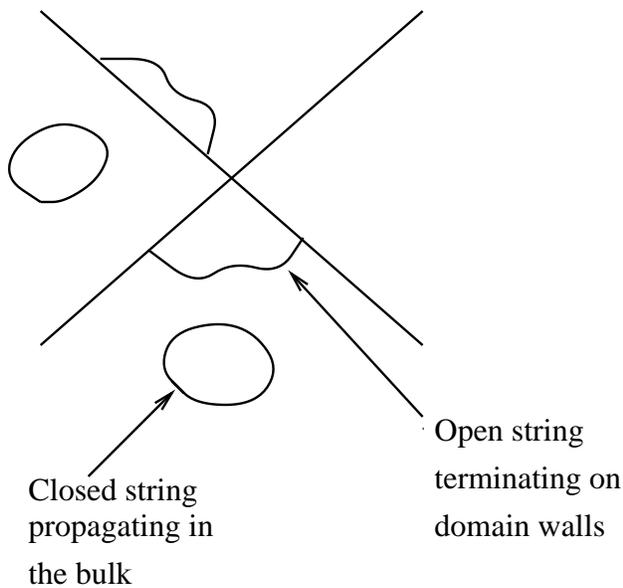}
\vspace{0.3in} 
\caption{{\small Open strings are attached to domain walls. }} 
\label{fig2}
\end{center}
\end{figure} 
\begin{eqnarray}
R~=~g_s~l_s~.
\label{radius}
\end{eqnarray}
In this case, the D4-branes discussed above are just the
M-theory fivebranes wrapped on the  
circle ${\bf S}^1$. Thus, all the branes in Fig. 3
are the M-theory fivebranes. All these fivebranes 
can be described as ${\bf R}^4 \times \Sigma$, where
${\bf R}^4$ stands for space of SYM  theory 
and   $\Sigma$ is a complex Riemann 
surface in a space with the following
complex variables: $v,~w$ and $t\equiv 
{\rm exp} (-s)={\rm exp} (-R^{-1}(x_6+ix_{10}))$. 
From the D-brane construction described above 
one finds that the curve $\Sigma$ is defined  by the equations
\cite {WittenMQCD}
\begin{eqnarray}
w &=& \zeta v^{-1} \nonumber \\[0.2cm]
w^N &=& \zeta^N t^{-1}~,
\label{sigma}
\end{eqnarray} 
where $\zeta$  is some complex constant. 
The discrete ${\bf Z}_N$ transformations  of SUSY gluodynamics
 are  realized as
\begin{eqnarray}
w~\rightarrow ~{\rm exp} \left( {i2\pi\over N}\right)~w,~~
t\rightarrow t,~~~v\rightarrow v.
\nonumber
\end{eqnarray}
The QCD strings discussed above are identified in this picture 
with the boundary of the  
M-theory membrane \cite {WittenMQCD}  
which intersects the corresponding fivebrane
\cite {Strominger}. 
\begin{figure}[h]
\begin{center}
\epsfbox{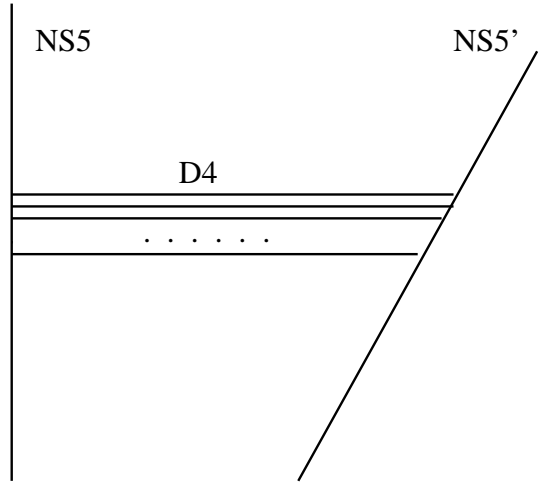}
\vspace{0.3in} 
\caption{{\small The D-brane setup.}} 
\label{fig3}
\end{center}
\end{figure} 
The M-theory membrane
lives in ${\bf R}^4\times Y\times {\bf R}$, where $Y$ is 
a complex three-fold which contains $\Sigma$. One can consider 
a membrane which is a product of a onebrane in ${\bf R}^4$
and a onebrane in $Y$. This membrane would look as a string to 
the four-dimensional observer living in ${\bf R}^4$.
Using this construction one identifies  both  the type IIA  and 
QCD strings (more precisely the strings which are in the same 
universality class). The type IIA string, from the M-theory
standpoint, would correspond to a membrane wrapped on 
the eleventh-dimensional circle ${\bf S}^1$. This gives  just
conventional type IIA string in the double dimensional reduction of 
M-theory \cite {Duff}. 
The tension of this string in terms of the M-theory 
variables is 
\begin{eqnarray}
T_{\rm IIA}~\sim~ {R\over l^3_M}~,
\label{TIIA}
\end{eqnarray}
where, $l_M$ denotes the fundamental length
of M-theory (inverse of the eleven-dimensional Planck scale).
 
Using the relations between the  M-theory 
parameters and type IIA parameters, 
$R=g_sl_s$ and $l_M^3=g_sl_s^3$, one finds that
$T_{\rm IIA}\sim 1/l_s^2$, as is expected for the fundamental type IIA 
string. 

Let us now turn to the QCD strings; their
existence  is more subtle to see. Consider an open curve 
$C$ in $Y$ parametrized by some parameter $0\le \sigma \le 1$.    
One chooses  $C$ in such a way that the endpoints of the curve are
in $\Sigma$ defined by Eq. (\ref {sigma}). Picking  one particular 
$C$ we determine a string in space-time. If there are $N$ 
different curves $C_k$ ($ k=0,...,~N-1$), they define $N$ strings.
$C_k$'s can be parametrized as follows:
\begin{eqnarray}
v& \propto& {\rm exp} (i2\pi \sigma/N) 
{\rm exp} (i 2\pi k/N),\nonumber\\[0.2cm]
 w &=& \zeta v^{-1}\, .
\label{Ck}
\end{eqnarray}
As a result, in this construction the QCD string tension is 
\cite {WittenMQCD}
\begin{eqnarray}
T_{\rm QCD}~\sim~{|\zeta|^{1/2} \over N l_M^3 }~.
\label{TQCD}
\end{eqnarray}
Since $\sum_{k=0}^{N-1} {\rm exp} (i2\pi k/N) =0$, 
the boundaries of  $C_k$ add up to zero. Thus, $N$ QCD strings
can join to create a closed loop in the complex 
$v$ plane which, subsequently, can be contracted with no 
obstruction, to a point. Thus, the QCD 
strings determined  above  can annihilate in groups of $N$ 
\cite {WittenMQCD}.

Let us now turn to the issue of how the BPS  domain walls of SYM
theory 
can be seen in the D-brane construction. If the domain wall 
interpolates between $x_3\rightarrow -\infty$
to $x_3\rightarrow +\infty$, it can  be described by a M-theory
fivebrane which interpolates between the fivebranes used to 
set  the vacuum states at  two ends. 
Thus, one can construct a  domain wall
as a fivebrane with the worldvolume $ {\bf R}^3\times S$,
where ${\bf R}^3$ is just the wall worldvolume $x_0,x_1,x_2$,
and $S$ is a certain three-surface embedded in the M-theory space-
time. 
When $x_3$ approaches  $-\infty$, $S$ looks as 
${\bf R} \times \Sigma$, with $\Sigma$ being defined by 
$w=\zeta v^{-1},~v^n=t$.
When $x_3 \rightarrow +\infty$, 
$S={\bf R}\times \Sigma^\prime$, where
$\Sigma^\prime $ is defined by
$w={\rm exp } (i2\pi/N) \zeta v^{-1},~v^n=t$. Thus, the 
brane interpolates between the neighboring 
chirally asymmetric vacua of the model. 
The tension of this wall can be also calculated \cite {WittenMQCD}, 
namely,
\begin{eqnarray}
T_{\rm D}\sim {R|\zeta| \over l_M^6}~.
\label{TDW}
\end{eqnarray}

The requirement for
the wall defined by ${\bf R}^3\times S$ to be a BPS state 
is equivalent to the condition that $S$ is a supersymmetric three-cycle
\cite {BeckerBecker}. 
Given these walls, it can be shown explicitly
that the open  QCD string which  
interpolates  from $v=v_0$ to $v=v_0 {\rm exp } (i2\pi/N)$ 
can actually terminate on the domain wall
\cite {WittenMQCD}. 
As a consequence, there should exist  
some degrees of freedom on the wall,  the open string endpoints, 
which transform in
the fundamental representation of the gauge group. 
These states are not present in the Lagrangian, but 
appear at the nonperturbative level in the theory. 
Since QCD  strings can annihilate in groups of $N$, 
this indicates that $N$ copies of 
fundamental representation can form a 
singlet ``baryon'' \cite {WittenMQCD}. 
The mass of this state  will scale as $N$. 

Could the properties of these states be understood 
within SYM theory or in the brane construction given above? 
In field theory this would be a difficult task, since it  requires 
solution  at strong coupling. 
One might hope that the problem is easier in the brane
construction. However, as we will argue below, 
the corresponding effective string coupling is large
in the regime we  deal with. Thus, the problem is 
equally complicated. 
The QCD string tension
is given in Eq. (\ref {TQCD}). On the other hand,  
$T_{\rm QCD}\sim \Lambda^2$. Using these two expressions  
one finds
\begin{eqnarray}
|\zeta|\sim N^2 \Lambda^4 l_M^6~.
\label{zeta}
\end{eqnarray}
Moreover, the domain wall tension (\ref {TDW}) 
must be proportional  to $N\Lambda^3$ \cite {DvaliShifman}.
Combining this 
assertion with (\ref {TDW}) and (\ref {zeta}), we obtain
\begin{eqnarray}
R\sim {1\over N \Lambda }~.
\label{R}
\end{eqnarray}
Finally, using the relation (\ref {radius}) we get
\begin{eqnarray}
N g_s\sim {1\over l_s\Lambda}~.
\label{gsls}
\end{eqnarray}

First of all, we see that 
 the string coupling scales as $g_s\sim 1/N$ in the large $N$ limit.
Moreover, since  $l^{-1}_s\gg\Lambda $, the combination  
$Ng_s \gg1$. This is the effective  coupling in the 
large $N$ limit. It is turns out to be large. Therefore, quantitative 
studies are not feasible \footnote{
Note that there is another reason why 
the brane construction cannot be used for 
quantitative studies. 
In addition to the D0-branes of type IIA,  which decouple
from SYM theory, 
there are Kaluza-Klein  states of the membrane worldvolume 
fields which are  wrapped $N$ times on ${\bf S}^1$. The masses
of these states scale as $1/(NR)\sim \Lambda$;    they do not decouple
from the SYM  degrees of freedom \cite {WittenMQCD}. That is why 
the model obtained in the D-brane construction is not SYM theory
{\em per se}, but
rather its close relative \cite {WittenMQCD}.} 
(for more detailed discussions of these and related issues
see Ref. \cite {HSZ}). 
Nevertheless, we can draw certain qualitative conclusions.

The result  most important for our purposes 
can be summarized as follows. The
QCD strings, as well
as type IIA strings, are described by the M-theory membranes.
A fivebrane describes 
the domain wall of SYM theory. The domain wall is a BPS state.  
This is guaranteed by the fact that the three-surface $S$ 
in the domain wall construction is a supersymmetric three-cycle.
As we mentioned above, 
the presence of this state in the theory is intrinsically
related to  the possibility of interpolation between 
distinct vacuum states of the theory.
Therefore, if one is to describe the
D-walls using  some  ``order parameter effective 
Lagrangian approach" in the large $N$ limit, the heavy states
should necessarily be included in  consideration
(despite  the fact that they are very heavy and, naively, 
should decouple). 
In other words, all conventional mesons are 
perturbative excitations of the closed QCD string.
If one writes down an effective Lagrangian including these fields only, 
one would  not be able to describe the BPS domain 
walls. To get such walls we  should necessarily include
a field (or fields) with mass  of  order ${\cal O}(N)$. 
Such a model Lagrangian was introduced in \cite {Gabad}, 
and used in \cite {DvaliGabadKakush} to study the BPS domain 
walls \footnote
{The reasoning for introducing 
a  new field  in \cite {Gabad} was  based on symmetry arguments.}.   

In Sec. 6 we discuss simple toy models engineered for the purpose of
 studying  the problem of interaction of the D-walls
and wall junctions with the light hadronic states.
Before we proceed to  these models, however, 
we will make a remark regarding non-supersymmetric Yang-Mills 
theory.

\section{Large $N$ (Non-Supersymmetric) Gluodynamics} 

If supersymmetry is broken, the  degeneracy of the vacuum states
inherent to SUSY gluodynamics is gone. Gone with it are
the (perfectly stable) walls. Recently it was argued, however 
\cite{EW,MS},
that in the large $N$ limit there exist infinitely many
quasistable ``vacua" [the decay rate of the false vacua was argued 
\cite{MS} to be suppressed as $\exp (-{\rm const}\, N^4)$];
the walls interpolating between them are quasistable. Their life time
is exponentially large, $\tau \sim \exp ({\rm const}\, N^4)$.
One may ask how the width of such walls scales with $N$.

The arguments of Ref. \cite{EW} were based on the D-brane picture;
 similar arguments were presented in Sec. 4. In Ref. \cite{MS}
a field-theoretic approach was exploited. One starts from
supersymmetric gluodynamics, adds the gluino mass to break SUSY, 
and then continuously interpolates from the limit of small masses
where reliable estimates are possible to the limit of large masses
when gluinos decouple, and one finds oneself in non-supersymmetric 
gluodynamics. If no phase transition happens {\em en route},
then the qualitative dependences obtained in the small mass limit 
persist in  non-supersymmetric 
gluodynamics.

Adopting the same approach and making the same assumption of
no phase transition in the problem at hand, we arrive at the conclusion
that in the large $N$ limit of non-supersymmetric 
gluodynamics the width of the (quasistable) domain walls 
scales as
$l\sim
(N\Lambda)^{-1}$. As in the supersymmetric case, this implies
that the dominant degrees of freedom comprising the wall
have masses $M\sim  N\Lambda$. In other words,
in the large $N$ limit there should exist some distinguished hadrons
with mass scaling as $N$ in pure Yang-Mills theory.

One can supplement the field-theoretic argument above by D-brane 
considerations which point in the same direction: 
the wall width in (non-SUSY)
gluodynamics scales as $1/N$.
Let us briefly review this construction. One starts with type IIA
string theory on ${\bf R}^4\times {\bf S}^1\times {\bf R}^5$.
To obtain the low-energy gauge theory one puts 
$N$ D4-branes on top of each other. 
The D4-brane worldvolume is taken to be 
${\bf R}^4\times{\bf S}^1$. The boundary conditions for fermions
on ${\bf S}^1$ are chosen to be antiperiodic. As a result, 
SUSY is broken on the worldvolume
and the low-energy theory is nothing but non-supersymmetric 
Yang-Mills theory with the gauge group U$(N)$. 

As follows from the results of Refs.
\cite {Mald,Pol,Wit,Wit1}, the large $N$ behavior of the SU$(N)$ 
part of this theory can be described by  string theory
on a certain  background $X$.
The topology of $X$ is ${\bf R}^4\times D\times {\bf S}^4$,
where $D$ is a two-dimensional disc. 
The metric of $X$ can be found explicitly \cite {Wit1}.
We just emphasize here the feature which is crucial 
for our discussion. 
The metric on $X$ depends on a certain parameter, let us call it 
$\eta$.  A (non-supersymmetric) gauge theory is expected to emerge 
\cite {Wit1} in the 
limit $\eta\rightarrow 0$. On the other hand, if $\eta\gg 1$,
one expects supergravity to be a good description of
the string theory on $X$. Thus, in order to use supergravity results
for studying the gauge theory, one has to assume that there is no
phase transition in $\eta$. Given this assumption, 
it is possible to  identify an object in string theory
which corresponds to a domain wall separating a given pair of 
distinct  vacua. This is a D6-brane which is
wrapped on the ${\bf S}^4$ factor of $X$ \cite {EW}.
As a result, one can easily establish the large $N$ scaling of the 
domain wall tension. The D6-brane tension is proportional to
$\sim 1/g_s l_s^7$. The string coupling $g_s $ scales as $1/N$.
Hence, the domain wall tension should scale as $\sim N$.
On the other hand, the volume energy density  generically scales as 
$N^2$.
In order to make the surface energy (tension) of the domain wall
scale as $N$ it is necessary to accept  that  its width 
is proportional to $1/N$. 

These arguments, combined with the field-theoretic arguments based on 
extrapolation from the supersymmetric limit, reinforce each other and 
indicate that 
there is no phase transition in 
the gluino mass.
Indeed, in the D-brane consideration one assumes
 that there is no 
phase transition in the parameter $\eta$
which, in general, has nothing to do with the gluino mass.
 
We conclude that  the scaling law for
 the non-supersymmetric walls is the same as in the supersymmetric 
case,
$l \sim 1/N$. The consequence which immediately follows is the 
existence 
of the ``solitonic glueballs'' which ``build'' the wall
and have  mass of order $N$.

The question as to the nature of these hadrons remains open. As 
far as we know, the wall-based consideration presented above
presents the first hint that such special hadrons may exist;
so far they have never been discussed in the literature. 

\section{Modeling the ${\bf Z}_N$ Vacua in the Large $N$ Limit}

As we discussed above, 
the vacua in SUSY gluodynamics  are defined by the 
vacuum expectation value
(VEV) of the order parameter 
$\lambda\lambda$. 
The effective Lagrangian for this order parameter
contains,  generally speaking,  an infinite number of massive fields. 
It is not known at present how  
to truncate self-consistently this Lagrangian.
However, in the large $N$ limit certain  simplifications
are expected to happen. Namely,  the
couplings of all physical states  
in the effective Lagrangian are expected to  be  suppressed by  powers
of $1/N$.  Thus, it seems that in the large $N$ limit one 
could concentrate
on the part of the  effective Lagrangian
which includes   $\lambda\lambda$ (and its SUSY partners)
only. 

In terms of the closed QCD string theory, this would correspond
to a certain truncated perturbative approximation to the string 
spectrum. 
However, as was discussed in the previous sections, 
this is  not  enough for the 
description of the D-walls of the model. 
As we established above, 
one should also include in the effective Lagrangian certain  
states (with mass of the order ${\cal O} (N)$) which have 
something to do with the ${\bf Z}_N$ structure of the ground state.
Let us study how this feature can actually be realized.
Below we will  work in the large $N$ limit, to the leading order in $1/N$. 
In fact, one appropriate toy model has been already discussed
in Sec. 3. Here we dwell on some other toy models with
the appropriate ${\bf Z}_N$ structure.

\subsection{A model ascending to the Veneziano-Yankielowicz 
Lagrangian}
 
The order parameter effective Lagrangian can be written  in terms of 
the chiral superfield \cite {VY}  
\begin{eqnarray}
 S\equiv \langle {\mbox{Tr}} (W_{\alpha}W^{\alpha}) \rangle=
 \langle {\mbox{Tr}} (\lambda^{\alpha}\lambda_{\alpha}) 
\rangle+\dots\equiv
 \langle \lambda\lambda\rangle+\dots~,
\nonumber
\end{eqnarray}
where $S$ is regarded as a classical superfield and the matrix elements 
above are defined in the presence of an appropriate coordinate-
dependent
background (super)source. 
The effective superpotential  reproducing all 
the anomalies of the model is given
by \cite {VY}
\begin{eqnarray}
{\cal W}_{VY}=NS \left[\ln\left({N\Lambda^3 \over S}\right)+1\right]~.
\label{vy}
\end{eqnarray}
As far as  dynamics of a single superfield $S$ 
is concerned, the superpotential (\ref {vy}) is (locally) exact. 
The corresponding scalar potential  
describes the spontaneous chiral symmetry breaking, with a 
nonzero gluino condensate \footnote{It also supports, 
for nonsingular K{\"a}hler potentials, 
 a chirally symmetric vacuum (the so-called Kovner-Shifman vacuum),
with the vanishing  gluino condensate
\cite {KovnerShifman}.} \cite {VY}. 

In terms of the closed QCD string theory, the expression 
(\ref {vy}) includes only a single  superfield of 
the closed QCD string perturbative spectrum.  
From this perspective it becomes clear that (\ref {vy}) should 
not be adequate for description of the D-walls,
since these require some nonperturbative stringy input as well. 

From the field-theoretical standpoint  this is reflected in the fact
that the  superpotential (\ref {vy}) does not respect 
the ${\bf Z}_N$ discrete symmetry  and should be modified 
\cite {KovnerShifman}. 
For the purpose of studying the vacuum structure {\em per se},
 the modification
by a constant integer-valued Lagrange multiplier is good enough \cite
{KovnerShifman}. However, to describe the domain walls, a  ``smoother''
modification is required \cite {KoganKovnerShifman}.  
This conclusion is supported by extensive analysis \cite{AVS,Smilga} of
the domain walls emerging in the Kovner-Shifman Lagrangian {\em per 
se}. 

One possibility to deal with this problem is to 
introduce an additional superfield $X$ which would 
restore the ${\bf Z}_N$ invariance of the model \cite {Gabad}. 
From the physical point of view, as we elucidated above, this would
correspond to some effective parametrization of the ${\bf Z}_N$ 
structure and the properties of the D-walls.
In general, 
it might be more relevant 
to introduce more than a single  superfield, let us say $X_1,X_2,...$,
to parametrize the strong coupling dynamics
of SYM theory. A new 
superpotential with multiple fields will be discussed in the next section.   
In the present section we 
concentrate on the case of a single superfield $X$ 
in order to elucidate how the construction works.
The ${\bf Z}_N$ symmetric  superpotential
with $S$ and $X$ fields 
can be regarded as the Veneziano-Yankielowicz superpotential 
(\ref{vy}) 
in which  the  
scale parameter $\Lambda^3$ is  promoted to some  ${X}$ dependent 
chiral
superfield, $\Lambda^3 N \rightarrow \Lambda^3 F(X)$ \cite{Gabad}.
Then,  dynamics of the ${X}$ field determines the phase of
the gluino  condensate. 
The  ${\bf Z}_N$ symmetric 
superpotential can be written as follows \cite {Gabad}:
\begin{equation}
 {\cal W}_{\small {\bf Z}_N}=
NS\left[ \ln \left(\Lambda^3 F(X)\over S\right)+1\right]~.
\label{SX}
\end{equation}
Here the function $F(X)$ can be presented as \footnote{For simplicity we  
use notations of \cite {DvaliGabadKakush} instead of those  of 
\cite {Gabad}.}
\begin{equation}
 F(X)=X\exp\left[{1\over N}\sum_{n} c_{n} 
\left(X\over N\right)^{nN}\right]~.
\end{equation}
The following properties are required:
\begin{equation}
 F^\prime (X^{\rm vac}_k)=0~,~~~F(X^{\rm vac}_k)=X^{\rm vac}_k\equiv
N\exp(2\pi ik/N)~.
\nonumber
\end{equation}
It is not difficult to see that the superpotential 
(\ref{SX}) respects the  ${\bf Z}_N$
discrete symmetry (with the transformations $S\rightarrow S\exp(2\pi
il/N)$, 
$X\rightarrow X\exp(2\pi il/N)$). The corresponding vacua are given by
\begin{equation}
 S^{\rm vac}=S_k= \Lambda^3 X^{\rm vac}_k~,~~~X^{\rm vac}=
N\exp(2\pi ik/N)~,
~~~k=1,2, \dots,N~.
\end{equation} 
Thus, the superpotential (\ref{SX}) can be used to describe the BPS
domain walls  interpolating between two distinct chirally asymmetric 
vacua at  large $N$.
Moreover, one can argue that, for the purpose of  description of the 
nearest-neighbor wall transition, 
the superpotential (\ref {SX}) can be simplified further
\cite {DvaliGabadKakush}. Indeed, 
the following
relation holds for the vacuum state labeled by the phase $k$: 
$$
S|_k=\Lambda^3
F(X)|_k\,. 
$$ This relation is nothing but the definition of the
gluino condensate with the phase set by the vacuum value
of the $X$ superfield.

 It is useful to introduce the chiral superfield
$S/F(X)$. In all  vacua this superfield takes the same value
equal to $\Lambda^3$. 
Let us now concentrate on interpolation 
between a pair of the nearest-neighbor vacua. In this case
the relative change in the gluino bilinear 
is of order $1/N$. 
Hence, the chiral
superfield $S/F(X)$ can only deviate from its vacuum value by
a quantity of order $1/N$. Thus, we  introduce 
the  parametrization
\begin{eqnarray}
{S\over F(X)}=\Lambda^3~\left(1- {\Sigma\over N}\right)~,
\label{sig}
\end{eqnarray}
with $\Sigma $ being a new chiral superfield.
Substituting Eq.
(\ref {sig}) in the superpotential
(\ref {SX}), one finds
\begin{eqnarray}
{\cal W}_{\small{\bf Z}_N}= N \Lambda^3 F(X) \left[ 1- {\Sigma^2\over 
N^2 }
+{\cal O} \left ( {1\over N^3} \right )  \right]~. 
\label{SigX}
\end{eqnarray}

The superfield $\Sigma$ enters this superpotential in the
subleading order in $1/N$.   
Thus, it is the superfield $X$  
which should describe the
domain walls between the adjacent vacua in the large $N$
limit. Neglecting higher-order terms, the superpotential then
is given by
\begin{eqnarray}
{\cal W}_{\small{\bf Z}_N}~=~ N \Lambda^3 F(X)~. 
\label{superX}
\end{eqnarray}
In addition, one should keep in mind that on a solution
the $S$ superfield is related \footnote{The subleading corrections in this
expression are  suppressed as $1/N^2$ since the $\Sigma$
field vanishes on the solution as $1/N$ 
\cite {DvaliGabadKakush}.} to the $X$ superfield as
follows:
\begin{eqnarray}
S=\Lambda^3 F(X) \Big (1+{\cal O} (1/N^2) \Big ).
\label{SvsX}
\end{eqnarray}
The superpotential (\ref {superX}) can be further reduced  
to the Landau-Ginzburg superpotential \cite {DvaliGabadKakush} 
\begin{eqnarray}
F(X) = X - {N\over N+1}~\left( {X\over N} \right)^{N+1}~.
\label{LandauGinzburg}
\end{eqnarray}

For this expression,
the BPS domain wall solution, with the width $l\sim1/N$,
can be explicitly found \cite {DvaliKakush,DvaliGabadKakush}. 
Indeed,  if one 
introduces a new variable $\Phi\equiv X /N$,
the corresponding BPS equation 
in the large $N$ limit takes the form
\begin{eqnarray}
\partial_z \Phi^*~=~i~\Big ( 1 - \Phi 
^N \Big )~. 
\label{LanGin}
\end{eqnarray}
This equation can be solved in the large $N$ limit.
Introducing the  notation
$$
\Phi \equiv \left(1-\frac{\sigma}{N}\right) e^{ i\tau/N}\,,
$$
where $\sigma$ and $\tau$ are two real
functions of $z$
with the boundary conditions $\sigma(\pm \infty) =0$ and
$\tau (-\infty) = 0,~~\tau (+\infty)=-2\pi$, 
one finds the solution \cite {DvaliKakush},
\begin{eqnarray}
{\rm cos} (\tau) &=& (1-\sigma ) {\rm exp} (\sigma), \nonumber 
\\[0.2cm]
\int_{\sigma(0)}^{\sigma(z)} dt \Big [ {\rm exp} (-2t) -(1-t)^2   
\Big ]^{-{1\over2}} &=& -N ~|z|~.
\label{solut}
\end{eqnarray}

The width of the wall is of order $1/N$. 
In this case, 
the presence of the D-wall in the theory is 
guaranteed by the presence of the field $X$ (which has 
mass ${\cal O}(N)$) in 
the superpotential (\ref {superX}). 
In reality, however, there 
might exist some light states  with masses of 
order ${\cal O}(1)$, which
would couple to $X$ with   couplings suppressed by the $1/N$
factors. 
The $S$ particle (or $\Sigma$) is  the example of such a state.
We would like to study the impact of this state
on the energy density and the width of the wall. 

\subsubsection{The energy density and  the wall width }

In this subsection we study how the presence of 
the $S$ (or $\Sigma$) field affects the energy density and 
the width of the BPS wall. Let us make a step back and consider
the superpotential (\ref {SX}) before  the   field $S$ (or $\Sigma$) is
eliminated. The 
tension of the wall is the sum of the kinetic and potential energy
contributions. The  BPS equations 
guarantee that these two contributions are equal. Therefore,
we can study only  the potential energy. The expression for the 
energy density of the wall  takes the form
\begin{eqnarray}
{\cal E} ~\propto~\int dz 
\left[
g_{XX^*}^{-1} \left|{\partial {\cal W}_{{\bf Z}_N} 
\over \partial X} \right|^2
+\left( g_{XS^*}^{-1} {\partial {\cal W}_{{\bf Z}_N} \over \partial X }
{\partial {\cal W^*}_{{\bf Z}_N} \over \partial S^*}+
{\rm h.c.}
\right)+ 
g_{SS^*}^{-1}
\left|{\partial {\cal W}_{{\bf Z}_N} \over \partial S} 
\right|^2~
\right].
\label{energy}
\end{eqnarray}

Generically, the K{\"a}hler potential is of order $N^2$. 
The fields $S$ and $X$ scale as $ N$. Hence, the 
K\"ahler metric is of order $N^0$ (more precisely, it cannot 
be larger than $N^0$). On the other hand, the derivatives of the
superpotential
with respect to $X$ and $S$ have a distinct large $N$ behavior. Indeed,
\begin{eqnarray}
{\partial {\cal W}_{{\bf Z}_N} \over \partial S }=
N \ln \left(\Lambda^3 F(X)\over S\right)~\sim N^0~,
\label{WS}
\end{eqnarray}
and, 
\begin{eqnarray}
{\partial {\cal W}_{{\bf Z}_N} \over \partial X }=
N S ~{F^{\prime}(X) \over F(X)}~\sim N~.
\label{WX}
\end{eqnarray}

We observe a very special pattern here.
The superpotential (\ref {SX}) scales as 
$N^2$, while the  $S$ field scales as $N$. Nevertheless, the 
corresponding derivative 
scales as $N^0$. This is  a consequence  of a very 
specific dependence
of the superpotential on the $S$ superfield.

From these results we find the contributions of each term
in the energy (\ref {energy}). The first term scales as 
$\sim N^2$, the second and third terms scale
as $N$ (at most), and the last term 
scales as $N^0$. All these terms are  to be
multiplied by the factor $1/N$. The latter  arises due to the integration 
over the wall width which scales as $1/N$.
Summarizing, the BPS wall tension scales as $N$, in accordance with 
\cite {DvaliShifman}. The dominant contribution to the tension is due to
the heavy $X$ superfield. The $S$ superfield can at most 
contribute to the tension at the level of $N^0$. 
This is negligible in the 
large $N$ limit. 
Therefore, in the leading order of the large $N$ expansion
the wall is made of the $X$ field. This is in full accord with the intuitive
expectations, of course.

Having  established this, let us analyze  
whether the interactions of the 
$X$ superfield with the $S$ (or $\Sigma$) field can cause 
the wall broadening.
As was mentioned before, the mass of these states
is of order $N^0$. If the walls were  able to emit these 
states, then the tails of the wall would  behave as  
${\rm exp} (-\Lambda z)$. This  would indicate that the wall
has a finite width, of order $1/\Lambda$, due to the ``cloud''
of finite mass states emitted by the wall. 
However, as we will show below,
the wall in our model cannot emit a 
finite number of light  states in the large 
$N$ limit. 

This assertion is related to the fact that the corresponding
couplings are suppressed 
as $1/N$. Indeed, let us consider the interaction  vertex  
of the $X$ superfield with the $\Sigma$ field (it is helpful to deal with 
$\Sigma$ rather than $S$). This vertex is defined as
\begin{eqnarray}
{\cal V}_{X^*\Sigma} \propto
\left.
 {\partial^2 \over \partial X^*
\partial \Sigma }~V (X,X^*,\Sigma,\Sigma^*)\right|_{\rm vac}~.
\label{vertex}
\end{eqnarray} 
Here $V$ stands for the potential of the model.
Using the superpotential (\ref {SigX}) we find that 
the vertex is proportional
to the off-diagonal element of the inverse K\"ahler metric
\begin{eqnarray}
{\cal V}_{X^*\Sigma} \propto
\left.~g^{-1}_{X^*\Sigma}\right|_{\rm vac}~.
\label{vertex1}
\end{eqnarray} 
For a generic form of the K\"ahler potential, 
the large $N$ scaling of this expression is not known. However, 
 in Ref. \cite {DvaliGabadKakush}
it was shown  that the 
wall solution of the theory with the superpotential 
(\ref {SX}) exists {\it if and only if} 
\begin{eqnarray}
\left.
g^{-1}_{X^*\Sigma}\right|_{\rm solution}~\propto~{1\over N}~.
\end{eqnarray}
Since the wall interpolates between the pair of distinct vacua, 
this implies that
$$
\left.
g^{-1}_{X^*\Sigma}\right|_{\rm vac}~\propto~{1/ N}\,. 
$$ 
Therefore, the coupling of the $\Sigma$ superfield to 
the wall is suppressed by the factor $1/N$. As a result,
the light states cannot be emitted by the wall in the 
large $N$ limit. The 
broadening of the wall will not happen, and 
the wall width will scale  as $1/N$. 

\subsubsection{Domain wall junctions} 

It has been  found recently that  in some models with the ${\bf Z}_N$
symmetry the domain walls
can form the so-called wall junctions which can be BPS-saturated
(preserve 1/4 of SUSY)
\cite{Townsend,Carroll} (see also \cite{TroitskyVoloshin}).
Such a  configuration is  depicted in Fig. 1 where we see that
$N$ domain walls join at some localized region
of space. The intersection of these walls forms a 
tube. 

Given that the model one  deals with 
has domain walls, it is natural to look for the junction type
solution as well. Let as assume  that the  $x$ and $y$
coordinates parametrize the plane of Fig. 1. Introduce
a complex variable 
$$
2\zeta \equiv x+iy\,.
$$
(Note the unconventional normalization.)
 In terms of this variable the 
corresponding BPS equations for the junction take the form
\cite{ChibisovShifman,Townsend,Carroll}
\begin{eqnarray}
\partial_{\zeta^*} X^*(\zeta)~=~
{\partial {\cal W}_{{\bf
Z}_N}\over \partial
X}~. 
\label{junctionBPS1}
\end{eqnarray}
Using the superpotential (\ref {superX}), 
(\ref{LandauGinzburg}) we obtain the 
 BPS equation for the junction,
\begin{eqnarray}
\partial_{\zeta^*} \Phi^*~=~1 - \Phi^N~.   
\label{junctionBPS2} 
\end{eqnarray}
Here, as above, we defined a new variable $\Phi=X/N$. A possible
constant phase in  the right-hand side of the BPS equation is 
absorbed in  $\zeta^*$.  

There are two possible solutions of the  junction type 
in the large $N$ limit. We will consider them in turn.
First, observe that 
if $|\Phi|<1$, the right-hand side of this equation equals to unity in the 
large $N$ limit. Thus, the solution of  the equation is
\begin{eqnarray}
\Phi = {\zeta },~~~{\rm for}~~~|\zeta|<1~.
\label{sol1}
\end{eqnarray}
If, on the other hand,  $|\Phi|=1$ then the solution of Eq.  (\ref 
{junctionBPS2})
is just a constant phase,
\begin{eqnarray}
\Phi = {\rm exp} (i\delta k)~, ~~~k=1,...,N,~~{\rm for}~~~
|\Phi|=1~.
\label{sol11}
\end{eqnarray}
Finally, if $|\Phi|>1$, the right-hand side of (\ref {junctionBPS2})
tends to infinity in the large $N$ limit. 
Therefore, the solution of this kind
does not exist. 

Summarizing, 
the solution we just found looks as the one presented in Fig. 1a.
There is a cylindrical core  of a unit radius  
in the center of the solution where $\Phi=\zeta$.
There are domain wall lines joining the cylinder 
from the exterior of the core. The value 
of the field $\Phi$ between these walls 
is  a constant phase ${\rm exp} (i\delta l),~l=1,2,\dots,N$,
which takes different values for different sectors 
of Fig. 1a. Notice that the value of the $X$ field in the center
of the solution (\ref {sol1}) is zero. If so, in accordance with  
(\ref{SvsX}), the value of the gluino condensate in this region 
 vanishes too. Thus, the (hypothetical) chirally symmetric 
Kovner-Shifman phase 
\cite {KovnerShifman}
is realized within this  solution inside the string 
at the geometric axis of the wall junction solution. 
We should also point out that the solution given by 
(\ref {sol1}) and  (\ref {sol2}) is defined up to  terms 
which provide  appropriate matching at the boundary circle
and at the boundaries of the sectors. 
The width of the region where 
the matching between (\ref {sol1}) and (\ref {sol11}) should be 
performed 
vanishes in the large $N$ limit and, therefore, these terms 
cannot be controlled in our approximation. Strictly speaking, the very 
fact
of matching remains an assumption.
 
Let us now turn to the second solution.
It is straightforward to check that  
the following expression is the solution of Eq.  (\ref {junctionBPS2}):
\begin{eqnarray}
\Phi~=~{\rm exp} \left ( {i 2\pi k \over N} \right
)\quad\mbox{in the $k$-th sector, }\quad k=1,2,....,N.
\label{sol2}
\end{eqnarray}
This configuration describes the domain walls intersecting
at the origin in the $(x,~y)$ plane (see Fig. 1b). 
The  intersection is just a straight line perpendicular to the $(x,~y)$ 
plane.  
Different sectors 
between different walls are parametrized by the  value of 
the phase of the solution (\ref {sol2}). 
Below, at the end of this section, we will argue 
that the two solutions presented above  
have the same energy in the limit of infinite $N$. 
 
Before we turn to this discussion, however,  let us calculate 
the tension of the tube which is formed at the intersection of 
the walls. This tension is defined by  circulation of the 
axial current along the big circle (call it $\Gamma$)
enclosing the tube \cite {GorskiShifman}
\begin{eqnarray}
T_{\rm Junction}~=~\int_{\Gamma} a_{l} dx^{l}~.
\label{jtension1}
\end{eqnarray}
In the model at hand $a_{\mu }=-iN^2(\Phi^*\partial_{\mu}\Phi~-~
\Phi \partial_{\mu} \Phi^*)/2$. Along the circle $\Gamma$ the axial 
current is a total derivative, $a_{\mu}=
N^2 \partial_{\mu} \alpha$,
where $\alpha$ is the phase which changes along the circle $\Gamma$.
As a result, the expression for the tube tension  is
\begin{eqnarray}
T_{\rm Junction}~=~2 \pi  N^2~.
\label{jtension2}
\end{eqnarray}

Thus, the tension scales as $N^2$ in the large $N$ limit. 
This is consistent with our previous estimate (\ref{dopone}) and with 
the
the expectation that there should be no  BPS strings
in SYM theory. The tension of the latter would scale 
\footnote{In the D-brane construction 
the tension of a BPS D-string
would scale as $1/g_s\sim N$. Since QCD 
strings annihilate in groups of $N$, they cannot possibly be 
BPS saturated objects.
The theorem  that in no non-Abelian theory  
BPS strings can exist  at weak
coupling is established in Ref. \cite{GorskiShifman}.
} as $N$. 

Let us now go back and show that  the two junction solutions
described above, although 
seemingly different, are both indeed BPS states
in the large $N$ limit.
Both of these solutions preserve 
$1/4$ of SUSY and satisfy the BPS equations. The 
only difference between them  
might be the energy of a configuration. 
However, simple counting based on  (\ref {jtension2})
and the expression for the BPS domain wall tension
$T_{\rm DW}$ \cite {DvaliShifman}, 
shows that the energies
of the two configurations enclosed by a circle of 
the radius $R$ are equal to $T_{\rm DW}N R+T_{\rm Junction}$. 
They
differ from one another only in the 
subleading order of the $1/N$ expansion. 
This may be interpreted as follows.
Suppose we start with the configuration of Fig. 1b
and consider the circle of a unit radius which encloses  the center.
As the number of walls tend to infinity, the angular separation between
the walls inside  the circle tends to zero. Thus, the interior
of this circle will look pretty much as the core in Fig. 1a. 
Hence, we have 
the same energy density in the region enclosed by the circle.
On the other hand, the angular separations between
the walls cannot be set equal to zero outside of the circle, 
since these walls  stretch out to the
spatial infinity along the radial directions.  

\subsection{Domain walls in a model with multiple fields}

In the previous subsection we found that the width of the 
BPS wall is not affected by interactions  with light 
gluinoballs as long as   these interactions 
are  suppressed as $1/N$. 
However, the story 
might be more complicated if there are open strings ending 
on the walls. 
In this case the wall is a sort of D-brane
for a non-critical SYM string. 
If these strings do exist, 
their lowest excitations  have nothing to do
with the conventional bound states, glueballs and gluinoballs.
Instead, they might appear as a result of the presence
of D-walls.
If so, 
the D-wall will be able to interact with states of the  open strings;
 this  interaction should not necessarily  be suppressed
by powers of $1/N$.  
 
The existence of the open strings 
cannot be  rigorously established in SUSY gluodynamics at present.   
Nevertheless, assuming that this  phenomenon
takes place, it is worth   finding a toy model in which these features
could be discussed.
Here we present a prototype field-theoretic 
model which has certain properties described above.

The model is specified  by allowing $F$
in Eqs.  (\ref {SX}),  (\ref {superX})  to be a function of $N+1$ variables,
\begin{eqnarray}
F~=~F(~X_1,X_2,\dots,X_{N+1})~.
\label{F}
\end{eqnarray}
After the $S$ field is removed, as described in the previous section,
the spectrum of the model  consists of  $N+1$ states.
One of these states is heavy, with mass of order $N$. 
The rest have finite masses. 
There is a BPS wall solution in the model. The wall is made of
the heavy field. The light fields vanish on the solution. 
The width of the solution is of order $1/N$, just like in 
the previous section. 
What is different   is that   the
interaction vertices  of the wall with the light states {\it are  
not} suppressed by $1/N$. 
Nevertheless, the  light degrees of freedom are not excited on the wall;
 there is no broadening. 

\subsubsection{The prototype superpotential}

Here we study  a ${\bf Z}_N$ symmetric model which
contains one heavy state and a number of light states.
The model is written in terms of $N+1$ fields denoted by
$X_k,~k=1,2,...,~N+1$. The ${\bf Z}_N$ symmetric superpotential
of the model is  (below we set $\Lambda=1$)
\begin{eqnarray}
{\cal W} ~=~X_1+e^{i\delta}X_2+e^{i2\delta}X_3+\dots
+e^{iN\delta}X_{N+1}~-~N~e^{-i\pi(N+1)-i\theta}~
\prod_{k=1}^{N+1}~{X_k\over N}~.
\label{prototype}
\end{eqnarray}
This superpotential replaces Eq. (\ref {superX})
after the $S$ field is eliminated and $F$ is allowed to be 
a function of $N+1$ variables. 
Here  $\delta~\equiv~2\pi/N$, and $\theta$ denotes the 
theta angle. 
One could have easily absorbed the factors $\exp(ik\delta)$
in Eq. (\ref{prototype}) in the definition of the
fields $X_k$. We keep them for reasons which 
need not concern us here. The model (\ref{prototype})
can be viewed as a simplified version of the model discussed
in Sec. 3.

The corresponding Lagrangian
is invariant under the simultaneous discrete ${\bf Z}_N$ transformations 
of the fields $X_k,~~k=1,2,\dots ,N+1$ 
\begin{eqnarray}
X_k~\rightarrow X_k~{\rm exp} \Big ( i\delta l\Big ),~~~
l=1,2,\dots , N.
\label{discrete}
\end{eqnarray}
In addition,  there is a symmetry which permutes different
$X$'s among themselves. A simplest pattern of these 
transformations can be written as follows:
\begin{eqnarray}
X_j~\rightarrow ~X_{j+1}~ {\rm exp} (i\delta), ~~~j=1,2,\dots ,N~,~~~
X_{N+1}\rightarrow X_1~.
\label{reshufling}
\end{eqnarray}

For simplicity let us chose the K\"ahler potential in the 
form
\begin{eqnarray}
{\cal K}~=~{1\over N}~\Big (X_1^*X_1+X_2^*X_2+\dots +
X_{N+1}^*X_{N+1}\Big )~.
\label{Kahler}
\end{eqnarray}
The reason why we introduce 
 the overall factor $1/N$  will
become clear shortly
(essentially, it is 
needed in order to make the K\"ahler potential scale as $N^2$
on the solution). 

Let us first study the vacuum structure in this 
model.
The vacua of the theory are determined by the  equations
\begin{eqnarray}
{\partial {\cal W}\over \partial X^j}=
\left[
e^{i\delta (j-1)} - {1\over X_j}N
e^{-i\pi(N+1)-i\theta}
\prod_{k=1}^{N+1}{X_k\over N} \right] =0\,,~~~
j=1,2,...,N+1.
\label{zeros}
\end{eqnarray}
This system of equations implies that
\begin{eqnarray}
X_1=e^{i\delta}X_2=e^{i2\delta}X_3=\dots
=e^{iN\delta}X_{N+1}~=~N~e^{-i\pi(N+1)-i\theta}
\prod_{k=1}^{N+1}~{X_k\over N}~.
\label{string1}
\end{eqnarray}
One finds as a solution 
\begin{eqnarray}
\langle X_k\rangle =N{\rm exp} \left( -i\delta (k-1) -
\frac{i\theta}{N}\right)\,, ~~~k=1,2,\dots , N+1.
\label{vacuum}
\end{eqnarray}
All other solutions of  Eq. (\ref {string1}) are related to the latter by the
symmetry transformations.
There are $N$ different vacua in the theory described by the 
superpotential
(\ref {prototype}), in full accordance with the (spontaneously broken)
${\bf Z}_N$. 

 Let us  comment on the role of 
the ``theta angle". When the $\theta$ parameter changes  
$$
\theta\rightarrow \theta +2\pi\,,
$$
 the vacuum values of different
fields transform into one another,
\begin{eqnarray}
X_1^{\rm vac} \rightarrow X_2^{\rm vac} \rightarrow 
X_3^{\rm vac} \rightarrow \dots X_N^{\rm vac} \rightarrow
X_{1}^{\rm vac}, ~~~X_{N+1}^{\rm vac} \rightarrow
X_2^{\rm vac}~. 
\label{secuence}
\end{eqnarray} 
Thus, the $2\pi$ shift of the theta parameter leads  
to a relabeling of the vacuum states,
precisely  as it should be  in pure SYM theory
(with massless gluinos and a nonzero gluino condensate). 
After the role of the theta term is 
elucidated, we will set $\theta=0$ for simplicity.   

Next, we turn to the task of finding the BPS domain walls 
for the superpotential (\ref {prototype}). 
The BPS equations for the minimal wall, interpolating between 
the neighboring vacua $m$ and $m+1$ are
\begin{eqnarray}
\frac{1}{N}\,\partial_z X^*_j=e^{i\gamma}
{\partial {\cal W}\over \partial
X^j} = e^{i\gamma}\left[
e^{i\delta (j-1)} - {1\over X_j}N
\prod_{k=1}^{N+1}{X_k\over N} \right]\,,\quad j=1,2,....
\label{BPS}
\end{eqnarray}
where 
$$
\gamma = {\rm Arg} \Delta {\cal W}\, , 
$$
and  the following  boundary conditions are implied
$$
X_j(-\infty) 
\rightarrow \langle X_j\rangle_{m}\,, \quad X_j(\infty) 
\rightarrow \langle X_j\rangle_{m+1}\,. 
$$
In Eq. (\ref{BPS}) and below it is assumed for simplicity that $N$ is 
even.

It is easy to see that the {\em Ansatz} which goes
through the system of equations  (\ref {BPS}) is
\begin{eqnarray}
X \equiv X_1 = X_2 e^{i\delta }=X_3e^{i2\delta}=\dots 
=X_{N+1}e^{iN\delta}.
\label{solution1}
\end{eqnarray} 
On this {\em Ansatz} one finds
(introducing a new variable $\Phi\equiv X/N$) that 
the system  reduces
 to a single equation,
\begin{eqnarray}
\partial_z \Phi^*=i\left( 1 - \Phi 
^N \right)\,,
\label{LG}
\end{eqnarray}
(in  the large $N$ limit the phase $\gamma =\pi/2$ up to 
subleading terms).

This latter equation is identical to (\ref {LanGin}) and 
can be solved in the large $N$ limit, as 
 in Eq. (\ref {solut}).
Therefore, expressions  (\ref {solution1}) and (\ref {solut}) 
define  the BPS domain walls of the model (\ref {prototype}).
The width of the wall is of order $1/N$, in accordance with 
the discussion in the previous sections. 

It is interesting to understand  what  
physical excitations ``make" the wall. 
To answer this question  we must determine
the  mass eigenstates of the model at hand.
The fields
$X_k$ in the superpotential (\ref {prototype})
are not diagonal. Indeed, the corresponding 
$N+1$ by $N+1$ mass  matrix of fermions has the form
\begin{eqnarray}
{\cal M}_{kj} =\left. N\frac{\partial^2{\cal W}}{\partial X_k \partial 
X_j}\right|_{\rm
vac}\, = \left\{
\begin{array}{c}
0\,\,\,\mbox{if}\,\,\, k=j\,,\\
e^{i(k-j)\delta}\,\,\,\mbox{if}\,\,\, k\neq j
\end{array}
\right.
\label{massmatrix}
\end{eqnarray}
where the mass matrix is evaluated at the vacuum
where $\langle X_1/N\rangle = 1$; in all other vacua the results
are essentially the same.
  Note
that each entry in this matrix is ${\cal O}(1)$. Nevertheless, upon
diagonalization we find one eigenstate -- call it ${\cal Y}$ -- with mass 
$N$
\begin{eqnarray}
{\cal Y} ={1\over \sqrt{N+1}} \left(  
X_1+e^{i\delta}X_2+e^{i2\delta}X_3+\dots
+e^{iN\delta}X_{N+1} \right)\,.
\label{YN1}
\end{eqnarray}

Other  $N$ states  $Y_l,~l=1,2,\dots,N$
 can be expressed in terms of  $X$'s in a simple way,
they have the same form as the diagonal generators of SU($N$), e.g.
\begin{equation}
Y_1 = \frac{1}{\sqrt{2}}\left(X_1-e^{i\delta}X_2
\right)\, ,\quad Y_2 = \frac{1}{\sqrt{6}}\left(X_1+ e^{i\delta}X_2
-2 e^{i2\delta}X_3
\right)\,,
\end{equation}
and so on. They all have mass 1, i.e. are light.

\subsubsection{Why the broadening of the wall does not take place}

In this subsection we argue that although 
the wall interacts with the light states with unsuppressed couplings, 
nevertheless the broadening of the wall width does not take place.
 
It is important  to note  that all  light states  $Y_l$ 
have the vanishing VEV's in the  
vacua of the theory. 
Moreover, these fields 
are identically zero on the domain wall solution
defined by Eqs. (\ref {solution1}) and (\ref {solut}).
The only field  which actually ``makes" the BPS wall is
the heavy state ${\cal Y}$. Moreover, the superpotential,
being expressed in terms of ${\cal Y}$ and $Y_l$, contains no terms
linear in $Y_l$. 

Let us now briefly discuss interactions of the heavy field ${\cal Y}$
with the light fields $Y_l$. 
The expression for the interaction potential is rather
cumbersome, so we merely summarize below its basic features.
In the superpotential there are vertices
with $N-1$ heavy fields and two light fields,
$N-2$ heavy  and three light, and so on. In the large $N$ limit 
the vertices have the
structure
$$
\left(\frac{{\cal Y}}{\sqrt{N}} \right)^{N-1} \, \frac{ Y_l}{\sqrt{N}}
\, \frac{
Y_{l'}}{\sqrt{N}}\,.
$$
The kinetic term is
$$
{\cal K} = \frac{1}{N}\, \left( {\cal Y}^*{\cal Y} +\sum_{l, =1}^N Y^{*}_l
Y_l
\right)\,.
$$
Rescaling the fields to cast the kinetic term 
in  the canonic form, one finds that, say,  the heavy-light-light vertex
 is not suppressed by  powers of $1/N$.
The production rate
is proportional to the square of the amplitude multiplied by
the inverse mass of the decaying state.
Since the decaying state has the mass of order $N$, 
this rate is going to be suppressed as $1/N$
$$
\Gamma ~ \propto
{\Lambda \over N}~.
$$ 
Likewise, one can find that the rates for the decays of the 
heavy state into arbitrary number
of light states are suppressed by the corresponding powers of $1/N$.  
Correspondingly, the wall energy is completely saturated by the 
contribution  of 
the heavy field ${\cal Y}$, the wall width $l\sim 1/N$,
and there is no broadening. 

\section{Discussion and Conclusions}

The vacuum structure of SUSY gluodynamics is rich 
and complicated. Essential ingredients in exploring
this structure are the  BPS domain walls  and junctions.
The theory is in the strong coupling regime.
Therefore, quantitative  studies of the fundamental model
are not feasible. Instead, one is able to abstract certain qualitative 
features
from exact results based on supersymmetry and from D-brane 
constructions.
It is also possible to
model essential properties of the vacuum 
in terms of effective Lagrangians. These model Lagrangians
allow one to explore the BPS
objects and to study the expected most salient features of the
theory. 

In particular, we argued that the minimal BPS wall in SUSY 
gluodynamics
 has width $(N\Lambda)^{-1}$ and  is ``made" of
the field which is not present among conventional mesons
of the model. The mass of the relevant state scales as $N\Lambda$. 

The width of the BPS wall
scales as $1/N$ despite
 the fact that the order parameter $\lambda\lambda$  interacts with
the  finite mass
mesons. This property allows one to naturally interpret
the tension of the  BPS wall  junction.
which  scales as 
$N^2$.  The $1/N$ scaling law for the width implies the existence of a
hadronic state with mass $\sim N\Lambda$ distinguished by its
 role in making the
wall. We suggested speculative
arguments hinting that such hadrons can persist in non-supersymmetric 
YM theory in the large $N$ limit. If so, there emerges a challenging task 
to
understand the nature of this special glueball.

\vspace{0.2in}

{\bf Acknowledgments}

\vspace{0.2in}

The authors are grateful to Gia Dvali, Sasha Gorsky, Dan Kabat, 
Zurab Kaku\-sha\-dze, Andrei Losev and Arkady Vainshtein 
for useful discussions.  A part of this work was done 
while  the authors  were visiting the
Aspen Center for Physics, within the framework of the program
{\em Phenomenology of Superparticles and Superbranes}.
We are  grateful to the Aspen Center for  Physics for the hospitality.
 The work of G.G. was 
supported by the grant NSF PHY-94-23002. The work of M.S.
was supported by  DOE under the grant number
DE-FG02-94ER408.


\end{document}